\documentstyle[preprint,prb,aps]{revtex}

\newcommand{\bmq}{{\mbox{\boldmath $q$}}}

\begin{document}
\preprint {WIS-99/14 Mar-DPP}
\draft
\date{\today}
\title{Extraction of nucleon momentum
distributions from  inclusive electron scattering on nuclei}
\author{A.S. Rinat and M.F. Taragin}
\address{Department of Particle Physics, Weizmann Institute of Science,
         Rehovot 76100, Israel}
\maketitle
\begin{abstract}

We   address   the   problem  of   extracting   single-nucleon   momentum
distributions $n(p)$  from inclusive  electron scattering data.   A model
for these relates nuclear and nucleon structure functions (SF) through an
intermediate SF $f^{PN}$  for a nucleus of  point-particles.  In addition
to the asymptotic limit (AL)  which depends on $n(p)$, $f^{PN}$ contains
generally  $q$-dependent Final  State Interactions  (FSI) parts.   In the
inverse problem one wishes to separate $q$-dependent FSI from the AL.  In
general  it suffices  to  know the  structure of  the theory,  but not
numerical results.   It appears,  that in the  $q$-range of  the analyzed
electron scattering
data,  FSI are  only weakly  $q$-dependent, making  it virtually
impossible  to  obtain parameters  in  a  free  fit of  the  parametrized
components of $f$.   Imposing a restriction, we obtain $n(p)$  for Fe and
$^4$He.

\end{abstract} \pacs{}

The simplest  bulk properties of a  many-body system in its  ground state
are the number (charge) and momentum distributions of the
constituents, or  of their  centers if  the latter are composite. Their
obvious   importance  invites   measurements, allowing the extraction of
these observables in  a model-independent fashion.  This appears
feasible  for  nuclear  charge  distributions, using  precision  data  on
elastic electron scattering and on muonic atoms \cite{chd}.
The situation is different  for single-nucleon momentum distribution (MD)
$n(p)$.  Contrary to the case  of charge densities, attempts to determine
$n(p)$  have   met  with difficulties,   which  are  frequently
circumvented by calling in theoretical MD \cite{cas,cc,dvn,ari}.

We   illustrate   the   above   for   exclusive   $A(e,e'p)(A-1)_n$   and
semi-inclusive $A(e,e'p)X$  reactions.  An  analysis usually  starts with
the Plane  Wave Impulse  Approximation (PWIA)  which expresses  yields in
terms of single-hole spectral functions.   For the two types of reactions
these  functions  are in  turn respectively linked to the  occupation
probability  $n_{\alpha}$ of  a  single-particle orbit  $\alpha$ and  the
total momentum  distribution $n(p)\,\,$ \cite{extra}. However, the PWIA
result is distorted by Final State interaction (FSI)
which  as a  rule  cannot  be removed  experimentally.   An
outstanding   exception   is   the   case   of   near-elastic,   high-$q$
electro-dissociation      of       the      $D$,      which     produced
$n^D(p)\,\,\,$\cite{bernh}.

A similar unsatisfactory situation  holds for totally inclusive processes
$A(e,e')X$, which may also be treated  in the PWIA and also require the
calculation of  FSI \cite{omar,cioffs}. Similar problems beset
indirect
attempts to  obtain MD.  These  are based on extracted  scaling functions
$\xi(q,y)$ which  depend on  the momentum transfer  $q$ and  some scaling
variable  $y$.    In  particular  Gersch-Rodriguez-Smith   (GRS)-type  of
theories \cite{grs} directly relate  $\lim_{q\to \infty}\xi(q,y\le 0)$ to
the  MD.    Plots  of  $\xi(q,y\le   0)$  as  function  of   $y$  display
coarse-grained scaling, i.e. clustering of  data points for different $q$
as function  of $y$.  Perfect  scaling is  interpreted as the  absence of
FSI, in principle enabling the  extraction of  $n(p)$, while  imperfect
scaling  manifests FSI  $\,\,$  \cite{sca,ciofs,rt,don}.   The latter  is
again difficult to remove experimentally.

In an alternatively approach one searches for a plateau in $\xi(q,\langle
y\rangle \le 0)$  for binned $\langle y\rangle$ as function  of $q$ or of
the squared  4-momentum $Q=q^2-\nu^2$  and which  is associated  with the
asymptotic   limit  (AL)   for   that  $y$,   and   thus  with   $n(y)\,$
\cite{ciofs,arr,rt2}.   Both   approaches  assume  that  FSI   terms  are
recognizable by a clear $Q^2$-signature.

We  now proceed  and present  a different  method which  proved eminently
succesful  in the  extraction of  MD from  data on  the SF  (or response)
$\phi(q,y)$ of liquid $^4$He$\,\,$\cite{gl,gl1,ri} and Ne \cite{az}.  For
systems with smooth  inter-particle interactions $V$, SF  may be expanded
in a series in $1/q$ \cite{grs}
\begin{mathletters}
\begin{eqnarray}
\label{a1}
\phi(q,y)=\sum_{n\ge 0}\bigg (\frac{M}{q}\bigg )^nF_n(y)
&=&\phi^{as}(y)+\phi^{odd}(q,y)+\Delta \phi^{even}(q,y)
\label{a1a}\\
F_0(y)=\lim_{q\to \infty}\phi(q,y)
&=&\frac{1}{4\pi^2}\int_{ y }^{\infty} dp p n(p),
\label{a1b}\\
\phi^{odd}(q,y)=\sum_{n,odd}\bigg (\frac{M}{q}\bigg  )^nF_n(y)
&=&  U^{(o)}(q)y\sum_{n}  a_n^{(o)}y^{2n}{\rm exp}(-[A^{(o)}y]^2)
\label{a1c}\\
\phi^{even}(q,y)=\sum_{n,even} \bigg (\frac{M}{q}\bigg )^nF_n(y)
&=& U^{(e)}(q)\sum_n a_n^{(e)}y^{2n}{\rm exp}(-[A^{(e)}y]^2)
\label{a1d}
\end{eqnarray}
\end{mathletters}
The above  decomposition is in  terms of coefficient  functions $F_n(y)$,
even and  odd in $y$, which  are preceded by factors with?,respectively,?
even and odd powers of $1/q$.  The even AL, Eq. (\ref{a1a}), depends in a
simple fashion on the MD.

Eqs. (1)  instruct how  to calculate  $\phi(q,y)$ from  $F_n(q,y)$, which
requires as  input $n(p)$, $V$  and special density  matrices \cite{grs}.
In  practice  this is  feasible  for  the AL  and  for  the dominant  FSI
coefficients $F_1(y),  F_2(y)$ \cite{ri}.  Alternatively one  may wish to
use SF data, in order to  conversely extract $n(p)$.  Eq. (\ref{a1}) then
serves  to express  information  on the  underlying  dynamics, which  one
incorporates   in  parametrizations   as  in   (\ref{a1c}),  (\ref{a1d}).
Algorithms show  how to  separate $q$-independent  AL from,  in principle
$q$-dependent,  FSI  without knowledge  of  computed  $F_n$.  From  fits,
including those for  the AL, one generates  a MD, which appears  to be in
good agreement with accurate calculations \cite{gl,gl1,ri}.

Two dynamical extensions complicate the above NR approach.  The first one
arises if  the bare inter-constituent  interaction is not smooth  or even
singular, which  requires the  replacement of  $V$ by  $V_{eff}=t_q$, the
$t$-matrix associated with $V$.  Since  $V_{eff}$ depends on $q$, the GRS
$q$-signature  of FSI  as in (\ref{a1c}), (\ref{a1d})  will be  modified,
although for quantum gases not drastically \cite{ri}.

A  second  complication  occurs  for molecular  systems  with  additional
degrees of  freedom, which may  be excited along  with the motion  of the
molecular centers of  mass.  Prime examples are  rotations and vibrations
in  a gas  of H$_2$  molecules.  Again,  because one  knows the  required
additional  dynamics  for NR  systems,  the  above complications  can  be
handled, both in calculations of the response and in methods to
extract $n(p)$.  As  long as FSI  parts possess  a distinct
$q$-dependence, the algorithms remain applicable \cite{ko,ital}.

We now turn to inclusive electron scattering from nuclei, for which the
cross section per nucleon reads
\begin{eqnarray}
\frac{d^2\sigma_{eA}(E;\theta,\nu)/A}{d\Omega\,d\nu}
=\frac{2}{A}
\sigma_M(E;\theta,\nu)\bigg\lbrack\frac {xM^2}{Q^2}F_2^A(x,Q^2)+
{\rm tan}^2(\theta/2)F_1^A(x,Q^2)\bigg\rbrack,
\label{a2}
\end{eqnarray}
The inclusive, as well as the  Mott cross section $\sigma_M$, are usually
measured  as functions of $E,\theta,\nu$ which are the beam energy,
scattering  angle  and  energy  loss.  The  above  SF  $F_{1,2}^A(x,Q^2)$
describe  the  scattering  of  unpolarized  electrons  from  non-oriented
targets  and are  conventionally expressed  as functions  of the  squared
4-momentum $Q^2=\bmq^2-\nu^2$ and the Bjorken variable $x=Q^2/2M\nu$ with
range $0\le  x\le A$.   The strong  variation in  both cross  sections in
(\ref{a2}) can be exploited by using the tempered, dimensionless ratio
\begin{eqnarray}
h_A(E;\nu,\theta)\equiv \frac {M}{2}\bigg
(\frac {d^2\sigma_{eA}(E;\nu,\theta)/A}{\sigma_M(E;\nu,\theta)}\bigg )
=\frac {xM^2}{Q^2}F_2^A(x,Q^2)+ {\rm tan}^2(\theta/2) F_1^A(x,Q^2)
\label{a3}
\end{eqnarray}
In  spite  of  technical  complications,   due  to  a  finite  number  of
constituents  which obey  Fermi  statistics, a  description of  inclusive
scattering of  relatively low-energy  leptons is not  basically different
from the above NR  case.  In particular one may still  venture to use the
notion of  a potential for the  description of $NN$ collisions  in FSI at
medium $q_{lab}\approx q \lesssim 0.5$  GeV. However, data for
the above kinematics yield only information on MD for restricted $p$.  In
order to  extend that range,  one needs considerably larger  momentum and
energy transfers $q,\nu$.  Those are provided by multi-GeV beam energies,
which may excite sub-nucleonic degrees of freedom.

Contrary to  NR systems,  there is  no way to  accurately compute  SF for
nuclei  with composite  constituents.   Whatever the  approach, it  seems
evident that with nucleons as major constituents, one has to relate
nuclear
SF $F_k^A$  in (\ref{a2}) to  those of the  nucleon $F_k^N$,  which in
practice is the $p,n$-weighted $F_k^{p,n}$ close to $F_k^D/2$. We shall
use below  such a previously formulated relation,  which refers only
to  nucleonic  and  sub-nucleonic   degrees  of  freedom  \cite{gr}.   It
specifically  disregards  virtual cloud pions (see ref. \onlinecite{pion}
\footnote
[1]{Eq. (\ref{a4}) is  easily interpreted in terms  of momentum fractions
(MF): The MF of a quark in a nucleus is the product of MF's of a quark in
a   nucleon  and   of  a   nucleon   in  a   nucleus.   SF  are  the
probabilities of  their occurrence. The $Q^2 \to \infty$ limit  of a
MF  equals  the observable Bjorken variable $x=Q^2/2M\nu$. Eq. (\ref{a4})
conjectures a similar
relation for large finite $Q^2$. Similar expressions have
been derived in perturbation theory \cite{jaffe} with the nucleon off its
mass-shell. $F^N_k$ in (\ref{a4}) relates to free nucleons.}.
\begin{eqnarray}
F_k^A(x,Q^2)&=&\int_x^{A}
\frac{dz}{z^{2-k}}f^{PN}(z,Q^2) F_k^N\bigg(\frac{x}{z},Q^2\bigg )
\label{a4}
\end{eqnarray}
The above contains SF for a nucleon $F_k^N$ and $f^{PN}$ for a nucleus of
point-particles, which has  to be computed.  Its asymptotic limit
depends  on the MD of
point-nucleons (or  of the centers of  composite ones) and on  FSI parts,
which account  for the  distribution of  the energy-momentum  transfer to
several core nucleons through $NN$ collisions.  In the kinematic range of
interest those occur at relativistic momenta $q$, for which the notion of
a  local,  energy-independent potential  breaks  down.   Again one  needs
$V_{eff}=t(q)$, which in general is off-shell, yet can be parametrized in
terms of observable $NN$ scattering data, and which permits a calculation
of the analog  of the SF $\phi(q,y)$ \cite{rt,rt1}.  As  a first step one
replaces there  the NR  GRS-West scaling variable  by a  relativistic one
\cite{gur}
\begin{eqnarray}
y=\frac  {M\nu}{q}\bigg (1-\frac  {q^2}{2M\nu}\bigg )\to
y_G\approx  \frac{M\nu}{q}\bigg   (1-\frac  {Q^2}{2M\nu}\bigg)=  \frac{M}
{\sqrt{ 1+4M^2x^2/Q^2)}}(1-x)
\label{a5}
\end{eqnarray}
Next one establishes the following transition from the SF
$\phi(q,y_G)$ to a relativistic analog \cite{rt1}
\begin{eqnarray}
f^{PN}(x,Q^2)=MD(x,Q^2)\phi[q(x,Q^2),y_G(x,Q^2)],
\label{a6}
\end{eqnarray}
where $D$ is a kinematic factor, which guarantees proper normalization of
$f^{PN}\,\,$.  The theory \cite{rt,rt1} based on (\ref{a6}) accounts well
for the data \cite{arr,na3}

We  now consider  the inverse  problem, i.e.  the extraction  of MD  from
inclusive scattering data.  We choose the recent Fe data for $E=4.05$ GeV
\cite  {arr} and  older $^4$He  data for  $E\le 3.6$  GeV \cite{na3}  and
determine the  reduced cross sections  $h_A$, Eq. (\ref{a3}).   Its right
hand side is  expressed by means of the nuclear  SF (\ref{a4}), which the
nucleon    SF   $F_k^N$.     The    latter   can    be   decomposed    as
$F^N=F^{N,NI}(x,Q^2)+\delta(1-x)\tilde   F^{N,}(Q^2)$,   containing   the
nucleon inelastic (NI)  and nucleon elastic (NE) parts.   The former have
been parametrized \cite{amau,br}, whereas  the latter are combinations of
known static form factors.

Eq. ({\ref{a6}) relates $f^{PN}$  to $\phi(q,y_G)$, with parametrizations
(\ref{a1})  as in  the  NR case.   In particular  its  AL (\ref{a1b})  is
generated by a MD, for which we choose a sum  of two  centered  gaussians
\cite{koh}
\begin{eqnarray}
n(p,\gamma_k)=
\frac {n(0)}{1+\epsilon} \bigg({\rm exp}{[-(p/p_1)^2]}+\epsilon
\,\,{\rm exp}{[-(p/p_2)^2]} \bigg )
\label{a7}
\end{eqnarray}
In initial runs we found it impossible to obtain a free fit for the FSI
part and  decided to generate starting values for parameters, by fitting
the theoretical FSI to (\ref{a1c}), (\ref{a1d}). In the relevant
$q$-range $\phi^{th}(q,y_G)$ in (\ref{a1}) appears to have the following
striking features:

i) $\phi^{odd}$  is negligibly small, while  the remnant
even FSI part  $\Delta\phi^{even}$ may reach 35$\%$  of  $F_0(y_G)$
for  $|y_G|\approx  0.2$  GeV, and up to  15$\%$ for
$|y_G|\approx 0.4$ GeV.

ii) $\phi^{odd}$ and $\Delta \phi^{even}$ are only weakly $q$-dependent.

Both features are properties of an effective interaction of a diffractive
nature, for instance generated by  a nearly imaginary $V_{eff}=t_q$ which
interchanges the roles  of odd and even parts in  $y_G$.  In the relevant
range, its strength Im$\,t_q \propto\sigma_q^{tot}$ happens to  be hardly
$q$-dependent, which  spells difficulties  for the  separation of  the AL
from  FSI with  that property
\footnote[1] {Not  only $NN$  interactions
produce  a  diffractive  $\sigma_q^{el}$.   For  example,  the  atom-atom
interaction  in  liquid  $^4$He   has  a  strong,  short-range  repulsion
\cite{aziz}.   This causes  damped  oscillation  in $\sigma_q^{tot}$  for
increasing $q$.   However for  all, but the  largest relevant  $q$, there
still is  a discernible $q$-dependence \cite{felt}.   The latter property
and the  relative weakness of FSI,  enable the extraction of  the MD from
the  dominating  AL.  The  most  extreme  example  is that  of  hard-core
interactions, which produce constant  $\sigma_q^{tot}$.  At least part of
the FSI are strictly  $q$-independent\cite{wein}.  The above implies that
in   the  large   $Q^2$-plateau  of   a  plot   of  $\xi(q,y_G)$   nearly
$q$-independent FSI $'$contaminate$'$ the  AL there.}
Additional evidence
may be implicit in an analysis by Ciofi $et\, al\,$ \cite{atti} who found
that  extracted  $q$-dependent  scaling  functions can  be  fitted  by  a
$q$-independent parametrization.   This surely is compatible  with weakly
$q$-dependent FSI.

We return to attempts to extract $n(p)$ from
Fe  data, which we divided  in 10 $\langle
q\rangle$    bins.     The    lowest    two,    which    correspond    to
$\theta=15,23^{\circ}$, resist any fit.  This observation is in line with
the  validity   of  the  relation  (\ref{a4}), which  predicts
deterioration
with decreasing $Q^2$.  For increasing $Q^2$, data in a given
$\langle q\rangle$  bin often  produce excellent  fits for  $h_{Fe}$ with
relative  deviations  from  data,  rarely exceeding  5$\%$,  and  usually
staying   under  $2\%$!   Nevertheless,   the  corresponding   parameters
occasionally appear far from their starting values.

The failure of a free fit forces the imposition of a restriction, e.g. a
given central value  $n(0)$ in the neighbourhood of  the theoretical one.
Such a choice leads to a well-determined, reasonable value for $p_1$, the
width  of  the  dominant  gaussian  in  the  MD  (\ref{a7}).   In  contra
distinction, the width of the second gaussian appears strongly correlated
to  its relative  strength  $\epsilon$ and  the  data only  approximately
determine $\epsilon^{1/3}p_2$.   All fits  produce a smooth  $n(p)$ which
are  primarily  differentiated  by  their  extention,  i.e.  by  the  rms
momentum.   For  He we  could  not  reproduce  the small  starting  value
$\epsilon=0.003$ and just put it to 0.  Table I assembles starting values
and fitted parameters.  The shaded area  in Fig. 1 shows the extremes for
$n^{{\rm Fe}}(p)$, corresponding to the  entries in the Table.  The drawn
line represents  the MD  for the  given starting values.   Fig. 2  for He
presents only one fit.

We  summarize.  Encouraged  by  the succesful  extraction of  single-atom
momentum distributions  from cross  sections for inclusive  scattering of
neutrons from mono-atomic quantum gases,  we attempted such an extraction
fron  nuclei.   The feasibility  in  the  case  of  NR systems  rests  on
knowledge  of  the atom-atom  interaction  in  the  above cases,  and  of
additional  rotation-vibration   dynamics  in   the  case   of  di-atomic
molecules, etc.  This knowledge enables parametrizations of the structure
function, incorporating the  above information.  Fits to  data, enable a
separation of $q$-dependent FSI parts from the $q$-independent asymptotic
limit.   That  AL  has  a  simple  dependence  on  the  desired  momentum
distribution, which is extracted in a truly model-independent manner.

The nuclear  case is in  many respects much  more complex.  For  one, the
composite nature of nucleons cannot be accurately ?order? described as in
the  molecular  case.   Consequently,  one   has  to  invoke  some  model
dependence, e.g. the  one implicit in (\ref{a4}).  Moreover,  there is a
real stumbling block in the shape of FSI, which are barely $q$-dependent.
Consequently a free fit to the data  does not lead to fits, smooth in the
parameters.   Only  restrained  fits  determine  characteristics  on  the
momentum distributions.

\bigskip

The authors  acknowledge stimulating discussions with  many colleagues on
the subject matter.  Special thanks go to Tim Shoppa and in particular to
Byron Jennings, who participated in initial stages of this research.

\bigskip
\par


{\bf Figure captions}

{Fig. 1}
Momentum distribution  $n^{Fe}(p)$ from fits of $d^2\sigma\,\,$
\cite{arr} for parameters  in Table I. Shaded areas correspond to entries
in Table I with $m_i$ corresponding to different fixed $n(0)$. The
drawn curve is  for the starting values of the parameters \cite{koh}.

{Fig. 2}
Momentum distribution  $n^{He}(p)$ from fits of  $d^2\sigma\,\,$
\cite{na3} for parameters in Table I. The drawn curve is for the starting
values of the parameters \cite{koh}.

\begin{center}
{\bf Table I}
\vskip 1cm
\begin{tabular}{|c||c|c||c|}
\hline
                      &\multicolumn{2}{c||}{Fe}      &He    \\
\hline
Parameters  & Starting values &extracted     & extracted
      \\
\hline
\hline
                      &\multicolumn{2}{c||}{69.8}     &        \\
$n(0)$(in fm$^{-3}$)&\multicolumn{2}{c||}{76.8}  &  80.51 \\
                      &\multicolumn{2}{c||}{58.3}     &        \\
\hline
                            &        & 0.837$\pm$ 0.011
&            \\
$p_1 $ (in fm$^{-1}$)              & 0.802  & 0.813$\pm$ 0.009
& 0.821  \\
        &        & 0.797$\pm$ 0.041
&            \\
\hline
                            &       &0.406$\pm$0.060
&              \\
$\epsilon^{1/3}p_2$ (in fm$^{-1}$) & 0.432 &0.365$\pm$0.060
&               \\
                            &       &0.268$\pm$0.036
&              \\
\hline
                            &        &1.177$\pm$0.136
&             \\
$p_2 $(in fm$^{-1}$)              &  1.390 &1.371$\pm$0.206
&             \\
                            &        &1.235$\pm$0.211
&            \\
\hline
                            &        &0.041$\pm$0.020
               &                   \\
$\epsilon$                  &0.03    &0.022$\pm$0.009
& 0.0           \\
                            &        &0.020$\pm$0.010
&            \\
\hline
                            &          & 1.108 $\pm$0.035
&       \\
$\sqrt{\langle p^2\rangle_{n(p)}}$(in fm$^{-1})$
&1.105& 1.105$\pm$0.037
& 1.005  \\
                                     &     & 1.013$\pm$0.021
&  \\
\hline
\end{tabular}
\end{center}
\vskip 1cm

Extracted parameters  for $n(p)$ of  the form Eq. (\ref{a6}).   For three
fixed  values  of $n^{Fe}(0)$  are  entered  the  width of  the  dominant
Gaussian $p_1$,
the  combination $\epsilon^{1/3}p_2$  of the  relative strength
$\epsilon$ and width  of the second one, as well  as each separately, and
the  root mean  square momentum.   For  He we present only  the fit  for
$\epsilon$=0.

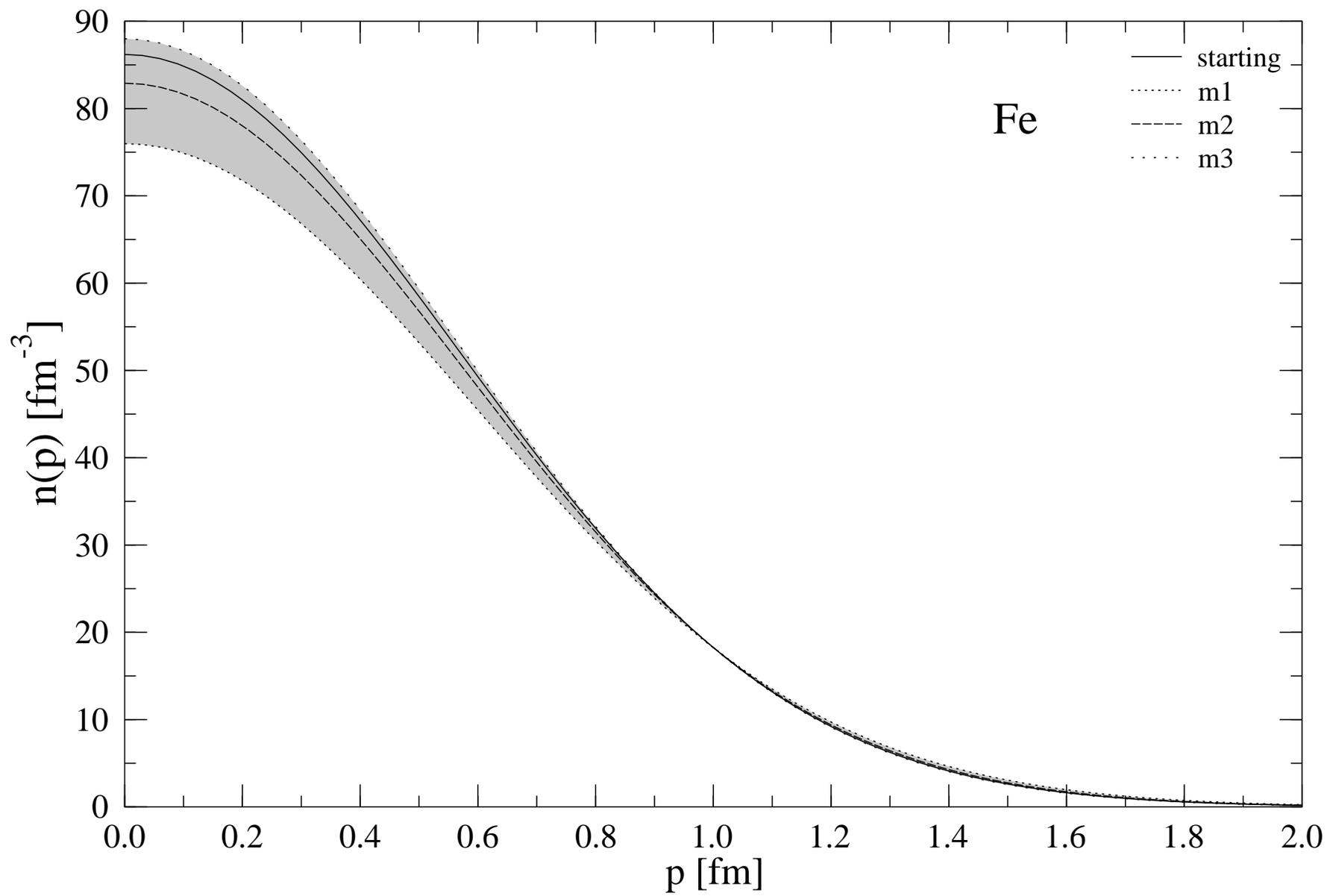
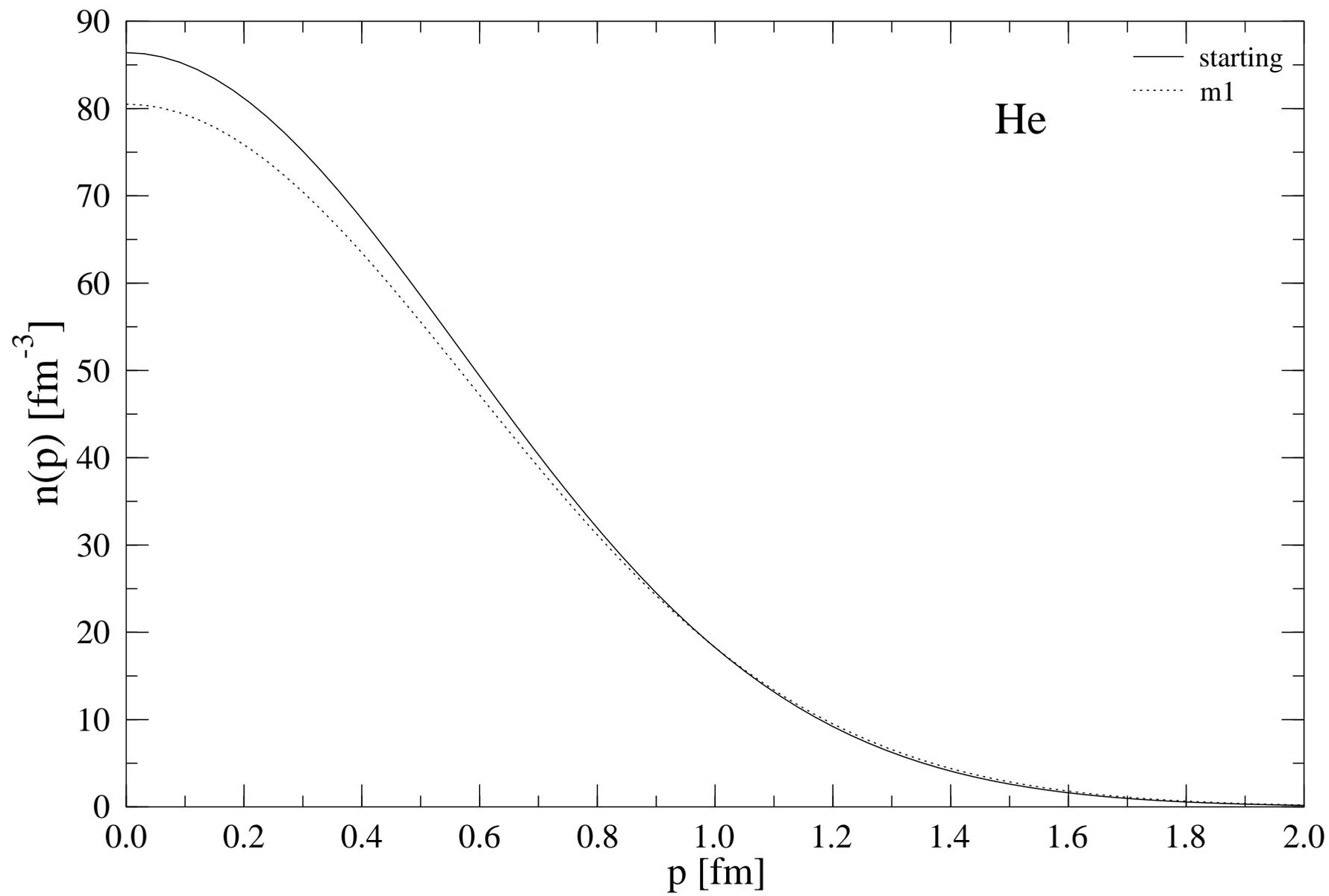
\end{document}